\documentclass[fdp,a4paper,fleqn%
]{w-art}
\usepackage{times,cite,w-thm}
\usepackage{amsmath}
\usepackage{mathptmx}
\usepackage{multirow}

\newcommand{\ttsmat}[4]{\big({ \textstyle {#1 \atop #3}{#2 \atop #4}}\big)}
\def\BNT{\,\hbox{\hbox to -0.2pt{\vrule height 6.5pt width .2pt\hss}\rm N}}
\def\BRT{\,\hbox{\hbox to -0.2pt{\vrule height 6.5pt width .2pt\hss}\rm R}}
\def\BZ{{\rm Z{\hbox to 3pt{\hss\rm Z}}}}
\def\IP{\relax{\rm I\kern-.18em P}}
\def\BCT{\,\hbox{\hbox to -3pt{\vrule height 6.5pt width .2pt\hss}\rm C}}
\def\BCS{\,\hbox{\hbox to -2.2pt{\vrule height 4.5pt width .2pt\hss}$
    \scriptstyle\rm C$}}
\def\BCSS{\,\hbox{\hbox to -2pt{\vrule height 3.3pt width
    .2pt\hss}$\scriptscriptstyle \rm C$}}
\def\BC{{\mathchoice{\BCT}{\BCT}{\BCS}{\BCSS}}}
\theoremstyle{plain}

\theoremstyle{definition}

\usepackage[]{graphicx}
\begin{document}
\DOIsuffix{theDOIsuffix}
\pagespan{1}{}



\hfill MPP-2010-21

\title[Basics of F-theory]{Basics of F-theory from the Type IIB Perspective}


\author[Ralph Blumenhagen]{Ralph Blumenhagen\inst{1,}%
}
\address[\inst{1}]{Max-Planck-Institut f\"ur Physik, F\"ohringer Ring 6, 80805 M\"unchen, Germany}
\begin{abstract}
These short lecture notes provide an introduction to some basic notions 
of F-theory with some special emphasis on its relation
to Type IIB orientifolds with O7/O3-planes.  

\end{abstract}
\maketitle                   





\section{Introduction}

Historically outstripping  heterotic string
compactifications and intersecting D-brane models,
the last two years  have seen the main
activity in the field of
string phenomenology shifting towards  F-theory
models. In this framework some of the 
model building
shortcomings  of D-brane realizations of grand unified 
theories (GUTs)  can be overcome.

To appreciate this, let us recall some issues on D-brane
constructions.
The gauge theories are supported on D-branes,
which in general can  have a dimension smaller than the bulk. 
Completely occupying  our observable large scale four-dimensional
world, they wrap certain sub-manifolds of the internal geometry.
The matter fields are localized on the intersections
of such D-branes. Since the early years of so-called intersecting
D-brane models, it was clear that this set-up  naturally allows
for semi-simple gauge groups with matter fields in the bifundamental
representations. Therefore, here one  directly engineers 
the $SU(3)_c\times SU(2)_w\times U(1)_Y$ MSSM, while leaving
the unification of gauge couplings at the GUT scale 
essentially unexplained. Indeed, the
gauge couplings depend on the generally different volumes of the 
internal cycles wrapped by the D-branes supporting each gauge factor.

It was quickly realized  that the construction of GUT groups 
$SO(10)$ and $SU(5)$ was obstructed by the perturbative absence 
of matter fields in the $16$ representation of $SO(10)$ 
respectively  
by the absence of the top Yukawa coupling ${\bf 10\,\, 10\,\, 5_H}$ 
for the $SU(5)$ case. These two latter features are of
non-perturbative origin for orientifold models 
(see \cite{Blumenhagen:2009qh} for a review). 
It was realized two  years ago  that the aforementioned problems 
with realizing  simple GUT groups in 
orientifold constructions are nicely
reconciled in F-theory models on elliptically fibered
Calabi-Yau
four-folds\cite{Donagi:2008ca,Beasley:2008dc,Beasley:2008kw,Donagi:2009ra}
(see \cite{Heckman:2010bq} for a more phenomenological review). 

One can think of F-theory as
Type IIB compactifications on compact complex n-dimensional 
manifolds  $B_n$ with 
general $(p,q)$ 7-branes wrapping $2(n-1)$ cycles  of $B_n$.
Since the 7-branes are of real co-dimension two, the solutions
to the Laplace equations are of logarithmic type. Therefore,
the backreaction of the 7-branes on the geometry and the dilaton 
is always substantial  and has to be taken into account. 
By  identifying the strong-weak $SL(2,\BZ )$ duality of the
Type IIB superstring with the modular group of a torus, 
C. Vafa \cite{Vafa:1996xn} 
showed that the backreaction can geometrically  be taken
into account by  an elliptic  fibration over the base $B_n$,
where the modular parameter of the fiber is identified
with the axio-dilaton field of Type IIB. The location
of the 7-branes correspond to the degeneration loci of the
elliptic fibration and  for supersymmetry the fibrations
have to be of Calabi-Yau type.

Due to the strong backreaction, only in a global $g_s\to 0$ limit
a general  F-theory model is expected to correspond to an 
orientifold.  F-theory inherently contains some features
which are non-perturbative from the orientifold point of view.  
This is the reason for the  appearance of exceptional groups 
in F-theory, which by a further breaking also realize
the spinor representation of a GUT $SO(10)$ as well as 
the top-quark Yukawa couplings ${\bf 10\,\, 10\,\, 5_H}$ in 
GUT $SU(5)$. For four-dimensional models, the basis $B_3$ is
a Fano three-fold and the 7-branes wrap complex surfaces, i.e. 
four-cycle.

Thus, F-theory is a non-perturbative completion of
Type IIB orientifolds where the 7-branes are completely
encoded  in the geometry of the elliptic fibration.
The aim of this lecture is to give an introduction
into some basis notions of F-theory, which essentially
addresses those students, who are already familiar with D-brane
constructions. It is explained why F-theory is inevitable
for the correct study of Type IIB compactifications with D7-branes,
in which sense it goes beyond the perturbative Type IIB  superstring
and how this leads to a  solution of the above mentioned problems with GUT
models.  Note that this was the first of a series of two lectures
on F-theory GUTs held  at the {\it 9th Hellenic School on Elementary Particle 
Physics and Gravity, Corfu 2009}.
The second lecture  focused more on the specifics of 
realizing four-dimensional $SU(5)$ GUTs from F-theory. 
Please consult  \cite{Denef:2008wq} for  more 
 detailed lecture notes on F-theory.


\section{D7-branes and $SL(2,\BZ)$ self-duality}
\label{sec_ftheory}

As a starting point, we consider Type IIB orientifolds
compactified on a Calabi-Yau three-fold $X$ and
an orientifold projection $\Omega \sigma (-1)^{F_L}$
(see \cite{Blumenhagen:2006ci} for a review).
Here $\sigma$ denotes a holomorphic involution of $X$
acting as
\begin{equation}
    \sigma^*(J)=J, \qquad   \sigma^*(\Omega_3)= -\Omega_3
\end{equation}
on the K\"ahler respectively holomorphic $(3,0)$-form
of $X$. This orientifold quotient
introduces an O7-plane into the theory, whose tadpole
is canceled by the introduction of stacks of D7-branes
wrapping various holomorphic four-cycles of $X$, whose
total homology class in $H_4(X,\BZ)$ is equal to the
one of the O7-plane. 
One can now compute the (chiral) massless
spectrum coming from the lowest excitations of 
open strings stretched between  various pairs of D7-branes.
This gives rise to gauge bosons of only unitary or
orthogonal/symplectic gauge groups and in addition
to matter fields transforming solely in bifundamental or
(anti)-symmetric representations of the gauge group.
This is simply a consequence of the fact that an open
string has two ends. Clearly, such open string excitations
can never give rise to exceptional  gauge groups and, as a group 
theoretic consequence, to matter in the spinor representation
of an $SO(10)$ gauge group. 
 
What we have just briefly described is the construction
of perturbative Type IIB orientifold string vacua and
its short-comings when it comes to GUT like structures.
Naively, this seems to be the end of the story.
However,   taking the perturbative string limit $g_s\ll 1$ is,
to  say the least, quite  questionable, if 
branes of (real) co-dimension two, such as D7-branes, are present.
This becomes evident by studying the D7-brane solution in Type 
IIB supergravity \cite{Greene:1989ya}, 
which is  magnetically charged under  the R-R  scalar field $C_0$ with
corresponding field strength $F_1=dC_0$. 

The space transverse to the D7-brane is two-dimensional so that
it is convenient to combine the two transverse coordinates
into a single complex variable $z=y^1+iy^2$. Furthermore ones combine the 
Type IIB dilaton together with the R-R scalar field $C_0$ into a complex
scalar field $\tau = C_0 + i e^{-\Phi}$. 
For the D7-brane supergravity solution, $\tau$ will be a function of the 
complex coordinate $z$, and the field
equation, i.e. the Laplace equation in two-dimensions, is now written as
$\partial_{\bar z}\tau(z,\bar z)=0$.
This means that $\tau$ must be a holomorphic function. 
However not any holomorphic solution is a good solution, e.g.
we must require ${\rm Im}\tau>0$. 
Furthermore the solution must have finite energy per unit volume, 
and it turns out that for this purpose one needs the $SL(2,\BZ)$ 
action on $\tau$. The solution can then be written as
$j(\tau)=z^{-1}$,
where 
\begin{equation}
  j(\tau)={\left(\vartheta_3^{8}(\tau) +\vartheta_4^{8}(\tau)+
    \vartheta_2^{8}(\tau) \right)^3 \over 8\, \eta^{24}(\tau)}=e^{-2\pi i\tau}
+ 744 + 196884 \, e^{2\pi i \tau} + \dots\    ,
\end{equation}
is  the modular invariant $j$-function.
Close to the D7-brane, i.e.  at $|z|\rightarrow 0$ this solution behaves as
\begin{equation}
\label{dilatcompl}
\tau(z)\sim{1\over 2\pi i}\log z  \;,
\end{equation}
which, circling once around the origin, gives rise to a 
monodromy $\tau\to \tau +1$. This monodromy reflects that the
D7-brane carries $C_0$ charge one. Due to this logarithmic dependence
of the axio-dilaton on the transverse coordinate $z$, the backreaction
of the D7-brane is so strong that one cannot really control the weak-coupling
regime. Once the string coupling is non-zero and maybe small somewhere,
it necessarily becomes large in other regions of the transverse space.
However, eq.\eqref{dilatcompl} implies that $g_s=\exp(\Phi)$ is  small 
close to the D7-brane  so that for the gauge theory
on the D7-brane we expect a weak coupling description.

As will now review, also the backreaction on the metric is strong. 
Consider the 10-dimensional space-time metric of the D7-brane solution:
${\rm d}s^2=-{\rm d}t^2+\sum_{i=1}^7{\rm d}x_i^2+e^{B(z,\bar z)}{\rm d}z\, {\rm
  d}\bar z$.
Then the Einstein  equation
connects the warp factor $B$ with the dilaton field, 
and one obtains the following simple relation
$\partial\bar\partial B=\partial\bar\partial\log({\rm Im}\tau)$.
It turns out that the solution with the correct modular 
properties and the right asymptotic behavior far away from the D7-brane
is given by
\begin{equation}
e^{B(z,\overline z)}=({\rm Im}\tau)\, {\eta^2(\tau)\, \bar\eta^2(\bar\tau)
 \over \left|\prod_{i=1}^N (z-z_i)\right|^{1\over 6}}\, .
\end{equation}
Here $N$ is the number of 7-branes and the $z_i$ denote 
their positions in the two-dimensional transverse space.
Expanding this function for large $|z|$ one gets that
$B(z,\overline z)\sim -N/12\cdot \log|z|$.
Using this asymptotic behavior one realizes that the metric goes like $|z^{-N/12}{\rm d}z|^2$ far away from the 7-branes.
This means that each 7-brane leaves a deficit angle of $2\pi/12$ in the 
transverse space.
In fact precisely 24 7-branes are required to get the deficit angle
$4\pi$ of the compact two-dimensional sphere $\,\BC\IP^1$.
Therefore, there exists a supersymmetric 
compactification of the Type IIB superstring
on $\,\BC\IP^1$  with precisely 24 7-branes and a varying
dilaton. 

So far we have only been talking about D7-branes.
However, due to the  $SL(2,\BZ)$ duality symmetry there
exist infinitely many different kinds of 7-branes.
Indeed, since there exists a doublet of  two-forms $(B_2, C_2)$  
in the ten-dimensional 
Type IIB string theory, there are not only fundamental
strings and D1-branes, but also strings carrying electric
charges $(p,q)$, where  in this notation
a fundamental string is a $(1,0)$ string. 
Acting with an $SL(2,\BZ)$ transformation
on the fundamental string gives
\begin{equation}
\label{pqsevenbranes}
              \left(\begin{matrix} p \\ q \end{matrix}\right)=
             \left(\begin{matrix} p & r \\ q & s \end{matrix}\right)
            \left(\begin{matrix} 1 \\ 0 \end{matrix}\right)\qquad
    {\rm with}\ ps-qr=1  \;. 
\end{equation}
Similarly, one has a doublet of ten-dimensional scalars $(C_0,\Phi)$
leading to magnetically charged $(p,q)$ 7-branes, 
where a D7-brane (charged only under $C_0$) is a $(1,0)$ 7-brane.
Since a fundamental string can end on a D7-brane, $SL(2,\BZ)$
implies that  a $(p,q)$ string can end on a $(p,q)$ 7-brane.
In eq. \eqref{dilatcompl} we have seen that the solitonic
solution of a D7-brane induces an $SL(2,\BZ)$ monodromy
$\smash{M_{\rm D7}=\ttsmat{1}{1}{0}{1}\Big.}$.
Applying the $SL(2,\BZ)$ symmetry, a general $(p,q)$ 7-brane
induces a monodromy
\begin{equation}
         M_{(p,q)}=   \left(\begin{matrix} p & r \\ q & s \end{matrix}\right) 
                      \left(\begin{matrix} 1 & 1 \\ 0 & 1 \end{matrix}\right)
                      \left(\begin{matrix} p & r \\ q & s
                        \end{matrix}\right)^{-1} =
                       \left(\begin{matrix} 1-pq & p^2 \\ -q^2 & 1+pq
                         \end{matrix}\right)\; .
\end{equation}
For later purpose we consider the three  7-branes
$A=(1,0)$, $B=(1,-1)$ and $C=(1,1)$.
It is now straightforward to compute the monodromy matrices
for the combinations of these three 7-branes 
listed    in table \ref{table_monoabc}.
\begin{table}[ht]
\begin{center}
\begin{tabular}{| c | c |c |}\hline
\ 7-branes\  &  number & monodromies   \\[2pt]\hline\hline
$A$ & $1$ & $M_A=\ttsmat{1}{1}{0}{1}\Big.$ \\
\hline
$B$ & $1$  & $M_B=\ttsmat{2}{1}{-1}{0}\Big.$ \\
\hline
$C$ & $1$  & $M_C=\ttsmat{0}{1}{-1}{2}\Big.$ \\
\hline
$A^n$ & $n$  & $M^n_A=\ttsmat{1}{n}{0}{1}\Big.$\\
\hline
$A B$ & $2$  & $M_A M_B=\ttsmat{1}{1}{-1}{0}\Big.$\\
\hline
$A^2 B$ & $3$  & $M^2_A M_B=\ttsmat{0}{1}{-1}{0}\Big.$\\
\hline
$A^2 B A $ & $4$  & $M^2_A M_B M_A=\ttsmat{0}{1}{-1}{-1}\Big.$\\
\hline
$A^n BC$ & $n+2$  & $M^n_A M_B M_C =\ttsmat{-1}{-n+4}{0}{-1}\Big.$ \\
\hline
$A^5 B C B $ & $8$  & $M^5_A M_B M_C M_B =\ttsmat{-1}{-1}{1}{0}\Big.$ \\
\hline
$A^6 B C B  $ & $9$  & $M^6_A M_B M_C M_B =\ttsmat{0}{-1}{1}{0}\Big.$\\
\hline
$A^6 B C B A $ & $10$  & \ $M^6_A M_B M_C M_B M_A =\ttsmat{0}{-1}{1}{1}\Big.$\  \\
\hline
\end{tabular}
\end{center}
\caption{Monodromies around stacks of 7-branes of types $A,B,C$.}
\label{table_monoabc}
\end{table}
To understand the massless modes between such more general $(p,q)$ 7-branes,
one notices that the $(p,q)$-strings can form so-called string
junctions. For instance a $(1,1)$ string can split into a
$(1,0)$ and $(0,1)$ string. 
Similar to open $(1,0)$ strings ending on D7-branes, for more general
7-brane there can exist so-called string junctions ending on them.
For instance there can be a string junction with  four external  strings
of type $(1,0)- (1,0) - (1,1)-(1,-1)$, which can end on the respective
7-branes $A- A - B -C$. Clearly, such objects can give
qualitatively new massless states beyond what is possible with perturbative
fundamental open strings.
We will come back to this in the next section.

\section{F-theory}

The  observations made in the previous section 
led C. Vafa in 1996 to the idea 
of F-theory. This is a hypothetical or rather auxiliary
twelve dimensional theory which, when compactified
on a two-dimensional torus, gives the Type IIB superstring.
The modular group of the torus is identified with the
$SL(2,\BZ )$ symmetry of Type IIB.
However, this twelve-dimensional interpretation 
is not meant in the sense of a standard Kaluza-Klein reduction,
as first there does not exist a twelve dimensional supergravity theory
with signature $(1,11)$ in the first place and second in ten dimensions 
there is no scalar field  corresponding to the
volume modulus of this $T^2$.  
Hence the 12-dimensional interpretation serves just to provide a geometrization
of the Type IIB  $SL(2,{\BZ})$ duality symmetry  rather than to
correspond to a real compactification from twelve to ten dimensions.
What makes  true sense though is to start with the eleven-dimensional
M-theory compactified on $T^2$ and define F-theory as the
${\rm vol}(T^2)\to 0$ limit.

The true power of this F-theory picture reveals itself
when compactifying the Type IIB superstring to lower dimensions.
We have just recalled that there should exist a compactification
of Type IIB on $\,\BC\IP^1$ with 24 7-branes preserving half the supersymmetry,
i.e. 16 supercharges. 
Observing that M-theory compactified on a $K3$ surface breaks half the
supersymmetry, one finds that F-theory compactified on an elliptically
fibered $K3$ with base $\,\BC\IP^1$ is 
the Type IIB string compactified on the   base $\,\BC\IP^1$ with 24 7-branes. 

For this to make sense, we have to find the 7-branes in this purely geometric
description, where we recall that in Type IIB there exist 
not just ordinary D7-branes but also these $(p,q)$ 7-branes introduced
in the previous section.
We have seen that close
to a D7-brane at position $u_1\in \BC\IP^1$ the complexified dilaton behaves
like $j(\tau)\simeq {1/(u-u_1)}$.
Now $\tau$ is really the modular parameter of a geometric
elliptic curve and it is known from mathematics that the $j$-function
naturally appears in this context. 
For this purpose we explicitly write the elliptic curve\index{elliptic curve}
as the hypersurface $\IP_{1,2,3}[6]$ in the homogeneous coordinates $(z,x,y)$. 
The fibration over the base $B=\BC\IP^1$ can  then be written as the
hypersurface constraint
\begin{equation}
\label{tateform}
        y^2 + a_1\, xyz + a_3\, y z^3 = 
        x^3 +  a_2\, x^2 z^2 + a_4\, x z^4
              + a_6\, z^6\; ,
\end{equation}
where the coefficients $a_n$ are homogeneous polynomials of degree $2n$
of the two homogeneous coordinates $(u_1, u_2)$ on $\,\BC\IP^1$.
More correctly stated, the $a_n$ are sections of $K_B^{-n}$,
where $K_B$ denotes the canonical bundle of the base $B=\BC\IP^1$.
Note that $z=0$ defines a section of the elliptic fibration, i.e.
the divisor $z=0$ is the base $\,\BC\IP^1$.
Completing the square and the cubic term, this so-called 
Tate form can be written in the so-called Weierstra\ss\  form\footnote{For
  base $\,\BC\IP^1$ the Weierstra\ss form is sufficient, but for
compactifications to six and four dimensions the Tate is very
convenient.}
\begin{equation}
         y^2  =  x^3 + f_4\,   x  z^4
              + g_6\,  z^6\; .
\end{equation}
To express $f_4$ and $g_6$ in terms of the $a_n$, it is convenient to introduce
the objects $b_2=a_1^2+4a_2$, $b_4=a_1\, a_3 +2 a_4$ and
$b_6=a_3^2 +4 a_6$ so that
\begin{equation}
       f_4={1\over 48}\left( 24\, b_4 - b_2^2\right), \qquad    
      g_6={1\over 864}\left( 216\, b_6  -36\,  b_4 b_2  +b_2^3\right)\; .
\end{equation}

Given the Weierstra\ss\ form with sections
$f_4$ and $g_6$, the complex structure $\tau$ of the elliptic
fiber over a point $(u_1,u_2)$ is implicitly given by  
\begin{equation}
\label{degfiber}
j(\tau)={ 4\, (24 f_4)^3\over 4 f_4^3 + 27 g_6^2} \, ,
\end{equation}
where indeed  the $j$-function\index{j-function} appears.
Now, the location of the 7-branes 
should be at the zeros of the denominator
\begin{eqnarray}
 \Delta=4 f_4^3 + 27 g_6^2 =
       -{1\over 4} b_2^2 (b_2\, b_6 - b_4^2) - 8 b_4^3 -27 b_6^2 +9 b_2 b_4
       b_6 \; . 
\end{eqnarray}
This is  the so-called
discriminant  of the elliptic fibration and its zeros are 
mathematically  precisely the points where the torus degenerates.  
Note that $\Delta$  is a polynomial of degree 24 in $(u_1,u_2)$, 
and thus has 24 zeros. 
These points  are  the positions of the 24
7-branes on $\,\BC\IP^1$.

So far we assumed that the discriminant has 24 different zeros.
However, when some of these zeros coincide the elliptic fibration
further degenerates, i.e. certain 2-cycles shrink to zero size.
In the M-theory
description, M2-branes wrapped on these shrunken 2-cycles provide
new massless states, which give rise non-abelian gauge symmetries.
In fact there exists a classification by Kodaira \cite{Kodaira} of the
different types such an elliptic fibration over $\,\BC\IP^1$
can degenerate. As shown in table \ref{tab:KodaireClassification}, 
this classification
is of the A-D-E type expected for singularities on $K3$ respectively
enhanced gauge symmetries. 

\begin{table}[ht]
    \centering
    \begin{tabular}{|r|r|c|c|c|c|l|c|}
      \hline
      ${\rm ord}(f)$ & ${\rm ord}(g)$   & ${\rm ord}(\Delta)$ & fiber      & singularity   & comp. & \ local geometry    & monod.\\  \hline\hline
      $\ge 0$\ \ \   & $\ge 0$\ \ \   & $0$    & I${}_0$    & smooth        & 1          &                & $\ttsmat{1}{0}{0}{1}\Big.$  \\ \hline
      $    0$\ \ \   & $    0$\ \ \   & $1$    & I${}_1$    & dbl.~point    & 1          & $y^2=x^2+z$    & $\ttsmat{1}{1}{0}{1}\Big.$  \\ \hline
      $    0$\ \ \   & $    0$\ \ \   & $n$    & I${}_n$    & A${}_{n-1}$   & $n$        & $y^2=x^2+z^n$  & $\ttsmat{1}{n}{0}{1}\Big.$  \\ \hline
      $\ge 1$\ \ \   & $    1$\ \ \   & $2$    & II         & cusp          & 1          &                & $\ttsmat{1}{1}{-1}{0}\Big.$  \\ \hline
      $\ge 1$\ \ \   & $\ge 2$\ \ \   & $3$    & III        & A${}_1$       & 2          & $y^2=x^2+z^2$  & $\ttsmat{0}{1}{-1}{0}\Big.$  \\ \hline
      $\ge 2$\ \ \   & $    2$\ \ \   & $4$    & IV         & A${}_2$       & 3          & $y^2=x^2+z^3$  & $\ttsmat{0}{1}{-1}{-1}\Big.$  \\ \hline
      $    2$\ \ \   & $    3$\ \ \   & $6$    & I${}_0^*$  & D${}_4$       & 5          & $y^2=x^2z+z^3$ & $\ttsmat{-1}{0}{0}{-1}\Big.$  \\ \hline
      $    2$\ \ \   & $\ge 3$\ \ \   &\multirow{2}{*}{$n+6$}  & \multirow{2}{*}{I${}_n^*$}  & \multirow{2}{*}{D${}_{n+4}$} & \multirow{2}{*}{$n+5$} & \multirow{2}{*}{$y^2=x^2z+z^{n+3}$} & \multirow{2}{*}{$\ttsmat{-1}{-n}{0}{-1}\Big.$} \\ \cline{1-2}
      $\ge 2$\ \ \   & $    3$\ \ \   &        &            &               &            &  & \\  \hline
      $\ge 3$\ \ \   & $    4$\ \ \   & $8$    & IV${}^*$   & $E_6$       & 7          & $y^2=x^3+z^4$  & $\ttsmat{-1}{-1}{1}{0}\Big.$  \\  \hline
      $    3$\ \ \   & $\ge 5$\ \ \   & $9$    & III${}^*$  & $E_7$     & 8          & $y^2=x^3+xz^3$ & $\ttsmat{0}{-1}{1}{0}\Big.$  \\  \hline
      $\ge 4$\ \ \   & $    5$\ \ \   & $10$   & II${}^*$   & $E_8$     & 9          & $y^2=x^3+z^5$  & $\ttsmat{0}{-1}{1}{1}\Big.$  \\  \hline
    \end{tabular}
   \vspace{0.2cm}
    \caption{The Kodaira classification of singular fibers in elliptic
      surfaces. The local geometry of the elliptic surface around such an A-D-E singularity is modeled in terms of coordinates $(x,y,z)\in\BC^3$. In the last column the elliptic monodromy of the singular fiber is given in terms of a $SL(2,\BZ)$-matrix. }
    \label{tab:KodaireClassification}
\end{table}

Note that in particular  the exceptional gauge groups 
$E_6$, $E_7$ and $E_8$  can be realized as enhanced gauge symmetries  
in F-theory.
Clearly they  cannot be realized  by fundamental open strings of the 
perturbative Type IIB string, i.e. not just with $(1,0)$ strings and
D7-branes.  These enhancements must involve more general $(p,q)$ seven
branes and the corresponding string-junctions between them.

To get an idea how this works, we  compare the geometric monodromy matrices 
in table \ref{tab:KodaireClassification} with those listed for stacks
of $A,B,C$ branes in table \ref{table_monoabc}.  It is evident 
that for the A-D-E series and the three fiber
types ${\rm II,III,IV}$  we have a perfect  match. Moreover, the number
of 7-branes is in all cases identical to the vanishing order 
of the discriminant.
One can also show \cite{Gaberdiel:1997ud} 
that the string junctions ending on the stacks
of branes provide precisely the massless states to fill
out the adjoint representation of A-D-E gauge groups.
Note that  the $D_n$ series is realized by the $A^n BC\ $ 7-branes, which
indicates that the $BC$ pair can be considered as the 
non-perturbative description of an $O7$ plane. 
Therefore, F-theory goes beyond the perturbative Type IIB
orientifolds in that it allows for general $(p,q)$ 7-branes
and their corresponding $(p,q)$-strings. 
It is precisely this more general structure which realizes
the exceptional gauge groups and as a consequence all their
group theoretic consequences, such as matter in spinor representations
of $SO(10)$ or the ${\bf 10\, 10\, 5_H}$ Yukawa coupling for
$SU(5)$ GUT models.

\section{F-theory compactifications and the Sen limit}

Finally, to connect to the second lecture on F-theory, 
let us briefly comment on  lower dimensional compactifications
of F-theory. Instead of fibering the torus over a complex one-dimensional
base, one can consider fibrations over surfaces $B_2$ or three-dimensional
bases $B_3$. Supersymmetry then implies that the total space
should either be a Calabi-Yau three-fold (for $B_2$) or Calabi-Yau four-fold
(for $B_3$). One can still write down a Weierstra\ss\ model, where
$f_4$ and $g_6$ are  sections of $K_B^{-4}$ and $K_B^{-6}$.
The zeros of the discriminant define complex co-dimension one curves in 
$B_2$ respectively
surfaces in $B_3$ and  give the location of 7-branes.
In these cases, it is more convenient to use the Tate form \eqref{tateform}
of the elliptic fibration, as there exists a refinement
of the Kodaira classification, the so-called Tate algorithm, 
which allows to determine the gauge group essentially
from the vanishing order of the discriminant and the
sections $a_n$ (see \cite{Bershadsky:1996nh} for more details).
Now it can however happen that the singularity enhances further
where these co-dimension one objects intersect. Similar to
intersecting D-branes, this is where additional matter fields are localized.

In the case of an F-theory  compactification on a smooth 
Calabi-Yau four-fold $Y$ 
a couple of new issues need to be considered.
First, one can show that chiral matter only arises
on the intersection curve between two 7-branes, if there exists 
a non-trivial $G_4$-form
background (M-theory point of view). 
Second, one finds a non-trivial 
D3-brane tadpole cancellation condition, which in this case reads
\begin{equation}
        N_{D3} +{1\over 2} \int_{Y} G_4\wedge G_4 = {\chi(Y)\over 24}
\end{equation}
where $\chi(Y)$ denotes the Euler characteristic of the smooth 
(appropriately resolved) four-fold $Y$.

One can define a limit in which the string coupling  goes
to zero almost everywhere on  the base. This is the so-called
Sen-limit \cite{Sen:1996vd},  defined by rescaling 
$a_3\to  \epsilon a_3, \ a_4= \epsilon a_4, \ a_6= \epsilon^2 a_6$ 
and sending $\epsilon\to 0$.       
In this parameterization  one finds
\begin{equation}
       f_4={1\over 48}\left( 24\,\epsilon\,  b_4 - b_2^2\right), \qquad    
      g_6={1\over 864}\left( 216\, \epsilon^2\,  b_6  -36\, \epsilon\,  b_4 b_2  +b_2^3\right)\; .
\end{equation}
so that the discriminant becomes
\begin{equation}
       \Delta=-{\epsilon^2 \over 4} b_2^2 (b_2 b_6 - b_4^2) + O(\epsilon^3)\; 
  \qquad \Rightarrow \quad
  \, j(\tau)\simeq {  b_2^4\over \epsilon^2 (b_2 b_6 - b_4^2) }\; .
\end{equation}
Therefore, for $\epsilon\to 0$ the Type IIB string coupling
constant $g_s$ goes to zero almost everywhere except on the
locus where $b_2$ vanishes. Studying the monodromies
one finds a D7-brane on the locus $(b_2 b_6 - b_4^2)=0$
and an O7-plane where $b_2=0$. Therefore, the Sen-limit 
defines the region  in the complex structure moduli space,
where F-theory is (almost everywhere) weakly coupled
and a perturbative Type IIB orientifold description
is justified.

\begin{acknowledgement}
I would like to thank Benjamin Jurke and Dieter L\"ust for useful comments
about the manuscript.
\end{acknowledgement}

\end{document}